\documentclass[aps,prl,twocolumn,letterpaper,showpacs]{revtex4}
\usepackage{boxedminipage}
\usepackage{lscape}
\usepackage{float}
\usepackage[ansinew]{inputenc}
\usepackage{epsfig}
\usepackage{subfigure}
\usepackage{graphicx}
\usepackage{amsmath}
\usepackage{amssymb}
\usepackage{amsthm,url}
\usepackage[french,english]{babel}  
\usepackage{epstopdf}
\usepackage{amsfonts}
\renewcommand{\d}[2]{\frac{#1}{#2}}

\newcommand{\barray}{\begin{eqnarray}}
\newcommand{\earray}{\end{eqnarray}}

\newcommand{\beq}{\begin{equation}}

\newcommand{\eeq}{\end{equation}}

\begin{document}
\selectlanguage{english}

\title{Machine learning for many-body physics: efficient solution of dynamical mean-field theory}
\author{Louis-Fran\c{c}ois Arsenault$^{1}$, O.~Anatole von Lilienfeld$^{2,3}$, and Andrew J. Millis$^{1}$}
\affiliation{$^1$ Department of Physics, Columbia University, New York, New York 10027, USA\\
$^{2}$ Institute of Physical Chemistry, Department of Chemistry, University of Basel, Klingelbergstrasse 80, CH-4056 Basel, Switzerland\\
$^{3}$ Argonne Leadership Computing Facility, Argonne National Laboratory, 9700 S. Cass Avenue, Lemont, IL 60439, USA}

\date{\today}

\begin{abstract}
Machine learning methods for solving the equations of dynamical mean-field theory are developed. The method is demonstrated on the three dimensional Hubbard model. The key technical issues are defining a mapping of an input function to an output function, and distinguishing metallic from insulating solutions. Both metallic and Mott insulator solutions can be predicted. The validity of the machine learning scheme is assessed by comparing predictions of full correlation functions, of quasi-particle weight and particle density to values directly computed. The results indicate that with modest further development, machine learning approach may be an attractive computational efficient option for real materials predictions for strongly correlated systems.
\end{abstract}


\maketitle

\hyphenation{Brillouin}

The quantum many-body problem of predicting properties of systems containing electrons or other fermionic entities has challenged physicists and chemists for decades. This is so because the minus sign associated with fermionic exchange creates a host of difficulties including long-ranged entanglement and a Monte-Carlo sign problem. The net effect is to place the generic fermion many-body problem in the class of problems whose full solution is exponentially hard. Although new developments such as matrix product and tensor network methods may provide solutions to ground-state properties with only power-law cost, search for efficient approximate methods to handle a wide range of phenomena at a wide range of temperatures remains a key goal of condensed matter physics and quantum chemistry.

In this work, we investigate the use of Machine Learning (ML) \cite{MLbook} to leverage existing results and provide an efficient approximate solution to a generic class of problems in quantum many-body physics. ML is in essence a way to use a database of known solutions to infer information about a new problem. In the condensed matter physics context it has  been used as an intermediate step in molecular dynamics calculations\cite{SumpterNoidNeuralNetworks1992,Neuralnetworks_Scheffler2004,Manzhos2006,Neuralnetworks_BehlerParrinello2007,bpkc2010,Zhenwei2015}, to predict density functionals (so far only in the 1D context)\cite{ML4Kieron2012}, to obtain transmission coefficients for electron transport~\cite{QuantumTransportML2014} and, very recently, to  predict the fermi level density of states of weakly correlated solids \cite{Schutt2014} and  find formation energies of materials \cite{Faber2015-1}. These applications relate to classical physics and to single-particle quantum mechanics. In the quantum chemistry context, ML has been successfully applied to predict energies and other scalar properties of  molecules \cite{HarvardCleanEnergyProject,RuppPRL2012,Montavon2013,Ramakrishnan2015-1,Dral2015-1,Lilienfeld2015-1,Ramakrishnan2015-2,Bereau2015-1,Ramakrishnan2015-3}. In addition, non-ML ideas from data science have been recently proposed as ways to help the solution of the non-equilibrium many-body problem \cite{Freericks2014}.

We propose to use machine learning methods to solve true quantum many-body problems. A technical issue arises: in many applications of machine learning, including most of the ones referred to above, the goal is to infer a scalar property (e.g. an energy) of a model specified by a modest number of scalar parameters. However, the generic quantum many-body problem is the solution of a functional equation  relating an input function (for example a bare electron Green's function) to an output function (e.g. an electron self-energy). Here we build on previous work \cite{Arsenault2014} which involved learning a function specified by a modest number of input parameters, to develop a formalism capable of solving the more general problem of mapping a function to a function. In independent contemporaneous work, questions related to learning  functions have been studied in the context of density functional theory where one seeks to learn the relation between a position-dependent charge density and an exchange-correlation potential\cite{Li2014,Vu2015}.

The context for our work is Dynamical Mean-Field Theory (DMFT)\cite{dmft2}, a widely used approximate method for determining the properties of materials with strong electronic correlations. DMFT approximates the solution to  an interacting fermion system in terms of the solution of  an auxiliary quantum impurity problem. The  impurity model, although simpler than the full problem, is still a quantum many-body problem. The impurity model is specified by a hybridization function $\Delta (\omega)$; the many-body physics by a local Green function $G(\omega)$ or  self-energy $\Sigma(\omega)$. The hybridization function itself is obtained from a self-consistency condition which involves  $\Sigma(\omega)$ and an initial band structure which encodes the chemistry and crystal structure of the material in question and may be parametrized as a  bare or initial hybridization function $\Delta^0(\omega)$.  In standard applications, the self-consistency condition is solved by iteration. One may imagine a ML process to solve the impurity model (relating $\Delta$ and $G/\Sigma$) or a ML process to solve the entire DMFT self-consistency loop (relating $\Delta^0$ and $\Delta^f,G^f/\Sigma^f$). In this paper, we only present results for the full solution of DMFT. Use of ML as impurity solver may also be valuable as an intermediate step, enabling the rapid construction of a database of solved problems in the real-materials context. Our formalism is general enough to apply to this possibility.

We test our methods using the Hubbard model defined on a three dimensional cubic lattice with first and second-neighbor hoppings, with Hamiltonian $H=\sum_{k\sigma}\left(\varepsilon_k-\mu\right)c^\dagger_{k\sigma}c_{k\sigma}+U\sum_i n_{i\uparrow}n_{i\downarrow}$. Here $\mu$ is the chemical potential and $\varepsilon_k = -2t\sum_{\alpha=1}^3\cos\left(k_{\alpha}\right) - 4t'\left[ \cos\left(k_{1}\right)\cos\left(k_{2}\right) + \cos\left(k_{1}\right)\cos\left(k_{3}\right) + \cos\left(k_{2}\right)\cos\left(k_{3}\right)\right]$.  The bare hybridization function is $\Delta^0(\omega)=\omega+\mu-\left(\sum_k\frac{1}{\omega-\varepsilon_k+\mu}\right)^{-1}$. We define energy units such that the full non-interacting bandwidth $W=2$ where $W = 12t$ if $|t'| \leq t/4$ and $W = 8t + 16|t'|$ if $|t'| > t/4$. Varying the ratio $t^\prime/t$ changes the structure in the density of states, in particular shifting the location of the density of states peaks relative to the band center (see Section~I of the supplemental material for examples).

\begin{figure}[htpb!]
  \begin{center}
  \includegraphics[scale=0.26]{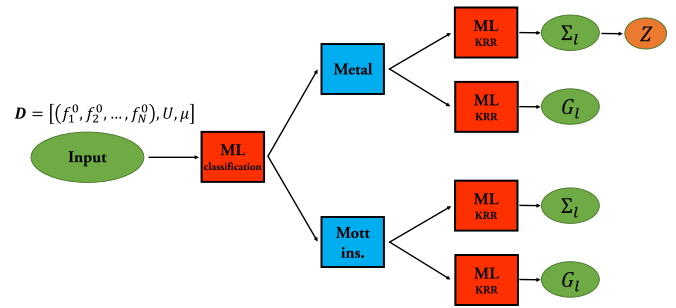}
  \end{center}
  \caption{(Color online) Schematic view of DMFT  as seen in a machine learning perspective. From an input description of a problem we are seeking a solution, the ML chooses first if the solution is metallic or insulating. Then the ML predicts the solution for the correlation function of choice by predicting the coefficients of the Legendre polynomials expansion of either the Green's function or self-energy. In the case of the self-energy for the metal, the ML predicted quasi-particle weight $Z$ can also be extracted.}
  \label{fig:ML_DMFT_scheme}
\end{figure}

We seek a machine that enables us to map a $\Delta^0$ to an output local Green function or self-energy. DMFT admits two classes of solutions: metallic ones with a non-vanishing density of states at the fermi level and a smooth self-energy, and Mott insulating solutions with a gap at the fermi level due to Coulomb repulsion and (in many cases) a self-energy with a pole near the chemical potential. We have found it advantageous to introduce a binary classification step that identifies a given solution as metallic or insulating and to use two different machines to determine the properties of the two kinds of solutions. For classification, we use the entire database minus one as training and the one remaining as the testing problem. We then repeat for all members of the database. We tested three different ML for classification: simple support vectors machine svm\cite{SVM} with $\sim 96\%$ accuracy, neural networks\cite{NeuralNetwork} with $\sim 97\%$ accuracy and decision forests\cite{DecisionForest} with $\sim 99.6\%$ accuracy. The only misplaced problems are critical metals extremely close to the transition. We only kept the decision forest as it outperformed the two others. Once the state of a new problem has been decided, the Kernel Ridge Regression (KRR) method \cite{MLbook,Arsenault2014} (more details follow) is employed to determine the solution using the sub-databases containing only metal or Mott insulating solutions. The full ML process for DMFT is shown in Fig.~\ref{fig:ML_DMFT_scheme} while some details about the parameters are explained later in the text.

The first step in implementing machine learning is to generate a database of initial conditions, in other words a set of bare hybridization functions that span a range of physically reasonable possibilities. We consider the set of hybridization functions defined by $t' = [0,-0.1t,-0.2t,-0.3t]$ (the case of positive $t^\prime/t$ could be accounted for by considering electron doping) and $\mu = 0$. Sections~I and II.A of the supplemental material give more details. We then obtain the database of solved problems by using the exact diagonalization (ED) method \cite{dmft2,Dagotto1994,Caffarel1994} to solve the single-site dynamical mean field approximation for interaction strengths in the range $0.16 \leq U \leq 4$ and densities in the range of $0.6 < n_d < 1.05$. Particularities of the ED database are discussed in Section~II.B of the supplemental material.

The second step in implementing machine learning is the construction of a representation of the information to be learned and of the {\em descriptor} $\mathbf{D}$, a unique identifier of a problem. Our input and output data are functions. Functions may be specified as a vector of coefficients in a space of basis functions $\phi_m$ (e.g. $\Sigma(z)=\sum_ms_m\phi_m(z)$). Our previous work\cite{Arsenault2014}, following work by Boehnke \emph{et al}.\cite{Boehnke_Legendre} found that Legendre polynomials were a very efficient choice of basis, so we adopt this representation here. The Legendre representation is most naturally formulated in imaginary time $0 < \tau < \beta$ with $\beta$ the inverse temperature and hence a correlation function is $f(\tau) = \sum_{l=0}^{\infty}\d{\sqrt{2l+1}}{\beta}f_lP_l(x(\tau))$\cite{Boehnke_Legendre}, where $P_l$ are the Legendre polynomials. The Fourier transform to $f(i\omega_n)$ can be done analytically\cite{Boehnke_Legendre}. The representation is general, we fit either the local Green's function or the self-energy as shown in Fig.~\ref{fig:ML_DMFT_scheme} (or even the hybridization function). See Section~IV of the Supplementary material for details.

The descriptor consists of the input function (hybridization function) plus a few scalar parameters; we denote the expansion coefficients of the function as $f$ and the scalar parameters $U$ (interaction strength) and $\mu$ (chemical potential) such that $\mathbf{D} = [(f_1,f_2,\ldots,f_N)_{input},U,\mu]$ (see Fig.~\ref{fig:ML_DMFT_scheme}). Note that both the full DMFT problem and the impurity solving part are the same problem as far as ML is concerned, the only difference being what database one chooses. The exact diagonalization method used here provides a representation of the input hybridization function in terms of bath level energies $\{\varepsilon_{ml}\}$ and hybridization parameters $\{V_{ml}\}$ ($m$ labels entries in the database and $l$ labels the different bath energies and hybridization parameters for a given entry in the database) so in practice we use these for the $f_m$. We have also implemented machine learning using the representation of the input function in terms of Legendre polynomials, with essentially identical results (see Section~V of the supplementary material). Section~II.A of the supplemental material shows how the bare ED parameters are obtained from a known band structure.

Machine learning then estimates the solution $f(z)\rightarrow \boldsymbol{f} = (f_1,f_2,\ldots,f_N)_{output}$ of a new problem in terms of an interpolation between known solutions. We use KRR, an expansion in the abstract multidimensional space of descriptors (each point $\mathbf{D}$ of this space represents a unique problem and the distance between two points is the distance metric), obtaining
\begin{equation}
\{f_m\}\approx \sum_{lm}\alpha_{lm}K_m\left(\mathbf{D}_l,\mathbf{D}\right),
\label{KRR}
\end{equation} where $l$ labels points in the dataset, $m$ labels entries in the output vector and the kernel $K$ is a function whose main characteristic is to weight most heavily the contributions of $l$ for which $\mathbf{D}_l$ is close to $\mathbf{D}$. As in \cite{Arsenault2014}, we use the weighted exponential kernel, and use the Manhattan distance between $\mathbf{D}_l$ and $\mathbf{D}$ (both are defined in Section~III.A of the supplemental material). The expansion coefficients $\alpha_{lm}$ are $\overline{\overline{ \boldsymbol{\alpha}}}=\left(\overline{\overline{ \boldsymbol{K}}}+\lambda\overline{\overline{ \boldsymbol{I}}}\right)^{-1}\overline{\overline{ \boldsymbol{f}}}$\cite{Arsenault2014}, where $\overline{\overline{ \boldsymbol{\alpha}}}$ is a matrix containing all the $\alpha_{lm}$, $\overline{\overline{ \boldsymbol{K}}}$ is the kernel matrix and $\lambda$ is a regularization parameter. $\lambda$ and the free parameter of the kernel are chosen using standard cross-validation, see Section~III.A of the supplemental material. In particular, as also found in \cite{Ramakrishnan2015-2}, we found that the actual value chosen for $\lambda$ is not really important. This formalism is very general and could be applied to the learning of other types of functions.
\begin{figure}[htpb!]
  \begin{center}
  \mbox{\includegraphics[scale=0.5]{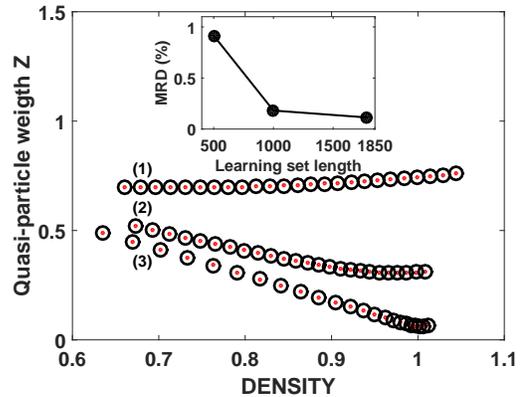}}
  \end{center}
  \caption{(Color online) Machine Learning predicted quasi-particle weight $Z$ (black circles) as compared to the exact results (red dots) as a function of filling of the impurity for different $U$ and $t'$  (1) $U = 0.64$ $t' = -0.3t$, (2) $U = 1.44$ $t' = -0.1t$, (3) $U = 2.08$ $t' = 0$. Inset: Median relative difference as a function of the size of the learning set}
  \label{fig:Z_ML}
\end{figure}
\begin{figure}[htpb!]
  \begin{center}
  \mbox{\includegraphics[scale=0.5]{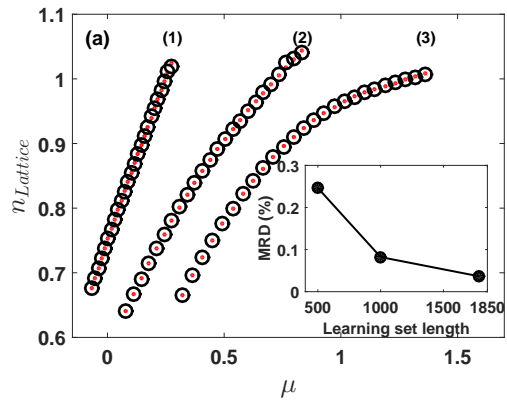}}
  \end{center}
  \caption{(Color online) Machine Learning predicted lattice density (black circles) as compared to the exact results (red dots) as a function of the chemical potential ($\mu$) for different $U$ and $t'$ (1) $U = 0.64$ $t' = -0.2t$, (2) $U = 1.44$ $t' = 0$, (3) $U = 2.08$ $t' = -0.1t$ (axis shifted $\mu+0.2$). Inset: Median relative difference as a function of the size of the learning set }
  \label{fig:density_ML}
\end{figure}

As first two tests of our predictive power we present scalar properties, the quasi-particle weight $Z=\left(1-d\Sigma'/d\omega|_{\omega\rightarrow 0}\right)^{-1}$ and the lattice density of electrons $n_{Lattice}=-2/\pi\int_{-\infty}^{0}\sum_k\text{Im}G_{Lattice}(k,\omega) = -2G_{Lattice}(\tau=\beta^-)$  as predicted from reconstructed correlation functions with ML obtained Legendre polynomial coefficients. We estimate $Z$ from a quadratic fit to the values of the reconstructed self-energies at the three lowest Matsubara frequencies(see for example \cite{Arsenault:Semon:Tremblay:2012} for why $Z$ can be estimated on imaginary axis). As easily seen from Fig.~11 of \cite{Boehnke_Legendre}, values of $G(i\omega_n)$ (or $\Sigma(i\omega_n)$) for the first few $\omega_n$ are given solely by the first few coefficients of the expansion in Legendre polynomials. Hence, the prediction of $Z$ shows how well the first few coefficients are learned. The results are shown in Fig.~\ref{fig:Z_ML} for typical values of interaction from weak to correlated metals and for different $t'$. The predictions for these specific $\mathbf{D}$ from the database are obtained by using all other examples as the training set. The predictions are in general very good with a slightly worst predicting power for larger correlation close to half-filling where $Z$ is close to zero. To study the error in a more rigorous way, we present in inset of Fig.~\ref{fig:Z_ML} what we call the median relative difference (MRD) for $Z$ as a function of the learning set size (see supplemental material Section~VI for details). This shows the median value of predictions for fifty different random examples each re-calculated with twenty different random learning set. As shown in the inset of Fig.~\ref{fig:Z_ML}, the MRD of $Z$ is slightly below $1\%$ for a small size of 500 and gets to around $0.1\%$ for the largest learning set. A predictive power of smaller than $1\%$ error even for a small database is very interesting especially since choosing completely random datasets is the worst case scenario.

In the case of the lattice density of electrons as a function of chemical potential, the ML path from Fig.~\ref{fig:ML_DMFT_scheme} is the one where we learn the $G_l$'s of the expansion of the local lattice Green's function then reconstruct it in imaginary time. Since $P_l(1) = 1$ for all $l$, the density is $n_{Lattice} = -\d{2}{\beta}\sum_{l=0}^{\infty}\sqrt{2l+1}G_l$. Therefore, contrary to the case of $Z$, the prediction of the density uses all predicted coefficients of the expansion. We show results in Fig.~\ref{fig:density_ML} for typical parameters, yet different than those presented for $Z$. To improve readability, we shifted curve (3) by 0.2. Once again the results are in good agreement with slightly worst predictions for $n_{Lattice} > 1$. This region tends also to be more problematic for $Z$. This is not fundamental but rather because our DMFT database is not as well constructed there. In the inset of Fig.~\ref{fig:density_ML} we show the MRD calculated the same way as for $Z$. ML does even better in this case where the MRD is at worst $\sim 0.25\%$.

\begin{figure}[htpb!]
  \begin{center}
  \mbox{\includegraphics[scale=0.5]{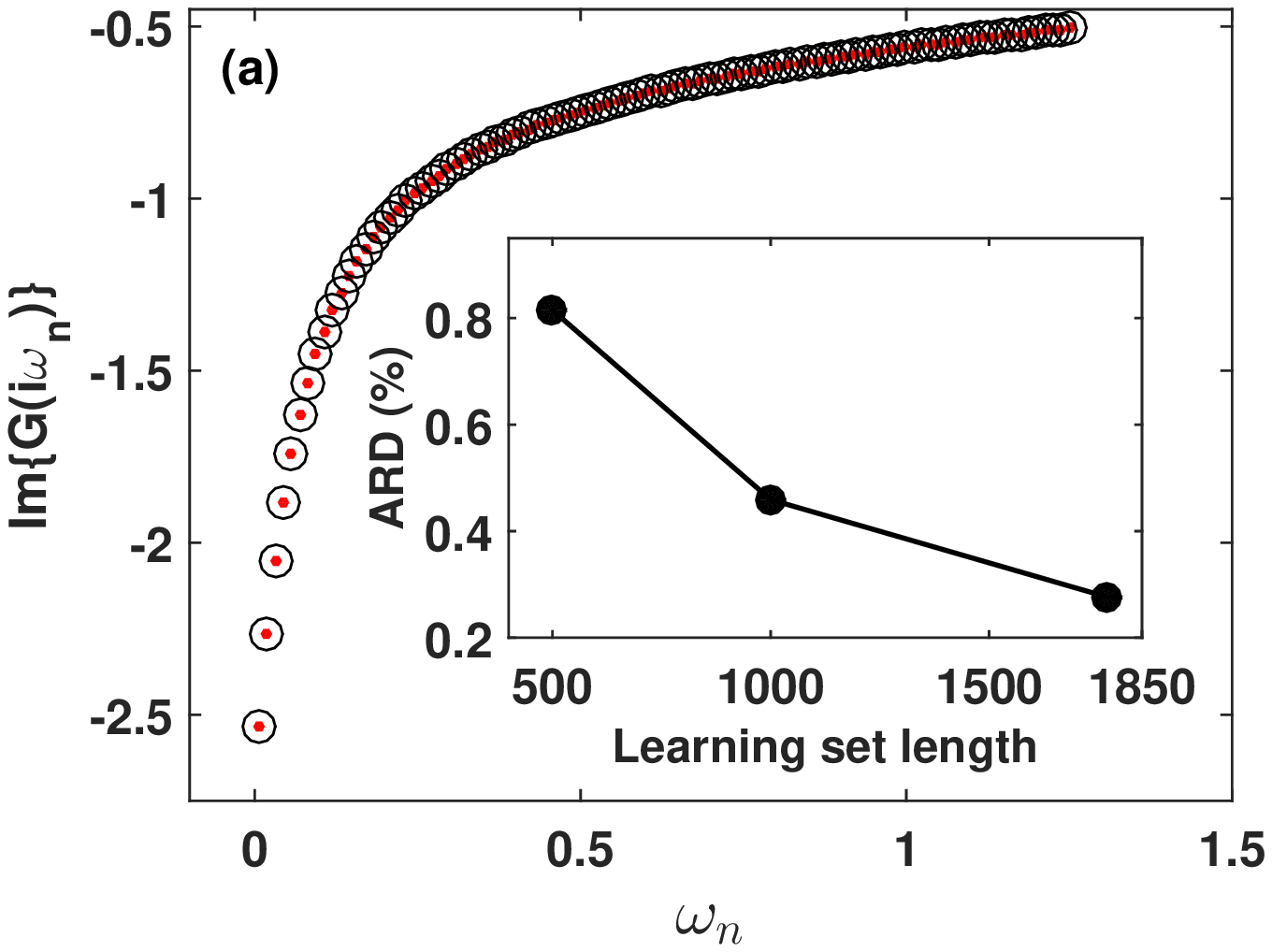}}
  \mbox{\includegraphics[scale=0.5]{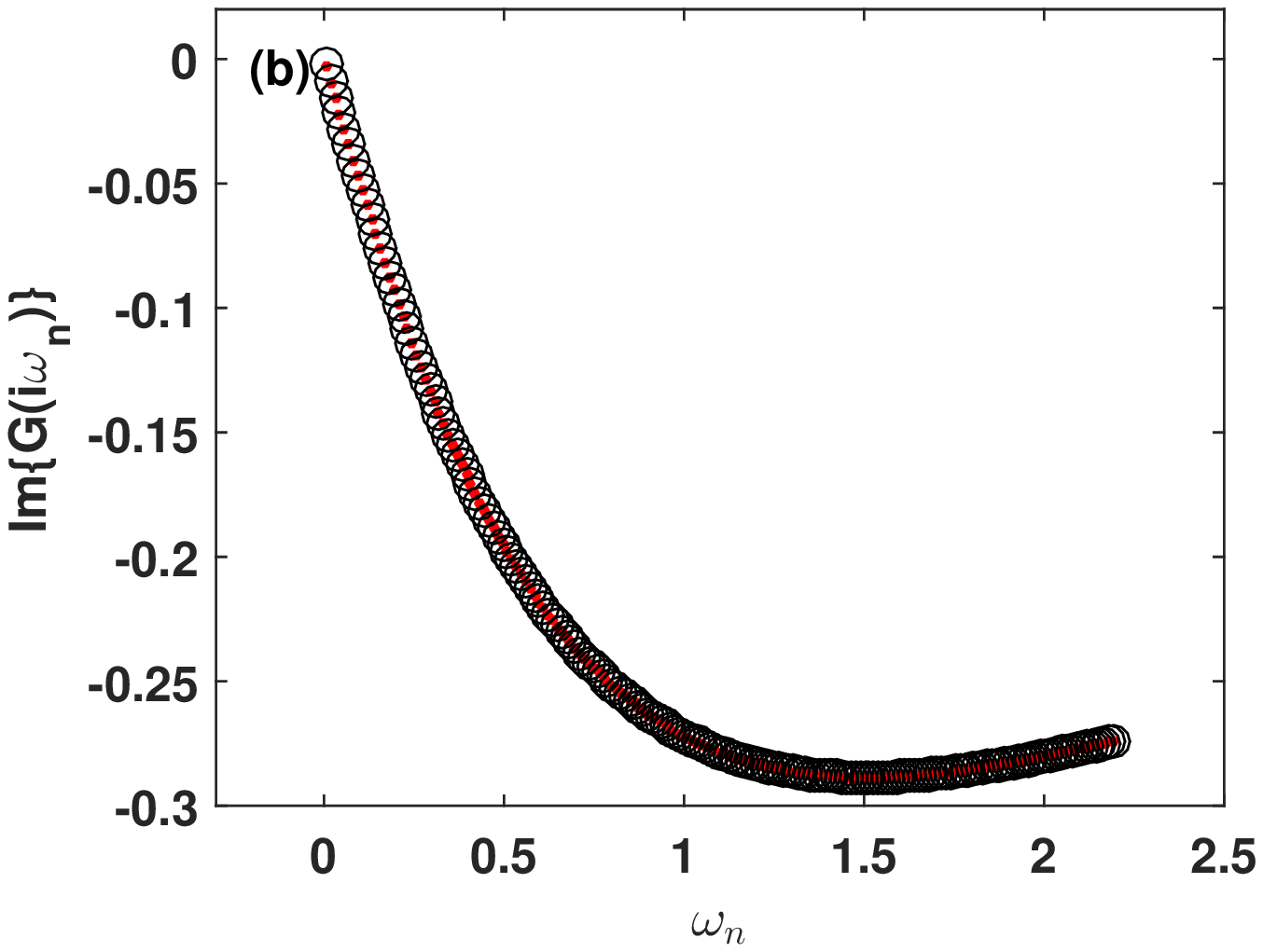}}
  \end{center}
  \caption{(Color online) Machine Learning predicted impurity Green's function (black circles) as compared to the exact results (red dots) (a) $U = 2.24$, $t' = 0$, $n_d \approx 0.92$, (b) $U = 3.68$ $t' = -0.2t$, $n_d = 1$. (c) ARD as a function of the size of the learning set for the metallic phase.}
  \label{fig:G_imp_wn}
\end{figure}
We now show in Fig.~\ref{fig:G_imp_wn}-(a) and -(b) the prediction of the imaginary part of the impurity Green's function in Matsubara frequency for two typical set of parameters. As can be seen, ML does a very good job at predicting both the metal and the insulator, although the number of insulating solutions in the database is not very large. In the inset of Fig.~\ref{fig:G_imp_wn}-(a), we present the average relative difference (ARD) for the metallic case as a function of the size of the learning set. The ARD was defined in \cite{Arsenault2014} as a way to measure on average the accuracy of the prediction of a full function using only one number. The values are obtained by averaging predictions for many random test sets. The global average prediction of a full function in the metallic case has an error in the worst case of $\sim 0.8\%$ which shows the predictive power of our ML scheme.
\begin{figure}[htpb!]
  \begin{center}
  \includegraphics[scale=0.5]{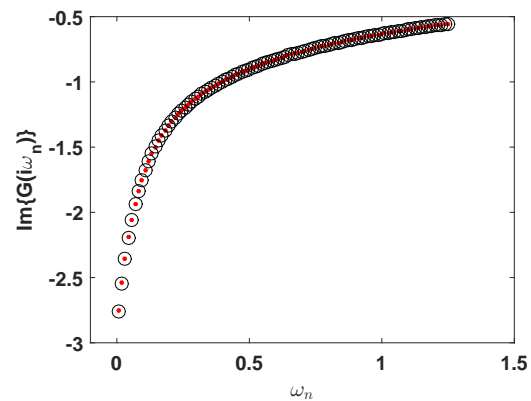}
  \end{center}
  \caption{(Color online) Machine Learning predicted imaginary part of the impurity Green's function (black circles) as compared to the exact results (red dots) for $U = 2$, $t' = -0.16t$, $n_d \approx 0.85$.}
  \label{fig:ML_G_predic_outside}
\end{figure}

We finally analyse the question of prediction of a totally new problem and the importance of training. Because our database is very homogeneous, for out of database predictions, we chose to use a width (arbitrarily set to be $5W=10$) larger than the actual lowest possible error in the cross-validation training used for previous results to avoid overfitting. In the supplemental material (Section~III.B.), we show how overfitting influences the predictive power of our ML approach. We show in Fig.~\ref{fig:ML_G_predic_outside} that indeed we can very well predict DMFT solution for new problems sharing no equal values of $U$, $t'$ and $\mu$ in the database by choosing as an example $t' = -0.16t$, $U = 2$ and $\mu$ such that $n_d\approx 0.85$. We trained a machine with the full database of 1783 metallic solutions, the metal being well predicted by decision tree classification. By this process (larger width), we loose some of the predictive precision we had for intra-database testing sets, but this is of no consequences as once we prove that we can accurately train a machine, what really matters is the predictive power for out of database unsolved problems.

In this paper, we have investigated how machine learning can be used in many-body physics as a method to predict correlation functions. We applied the scheme to DMFT and showed that we can accurately predict its solutions. Our approach maps input functions to output functions and can be applied without any changes as an impurity solver for DMFT rather than to learn the fully converged solution or for any other cluster embedding theory with a self-consistency relation. Impurity solving might be the best way to use ML for real materials predictions since accuracy depends largely on having a large database. It is also general enough to be applied to other problems where learning a function is important. The learning of a function using Kernel Ridge Regression  might be improved by adding a simple form of constraints in the minimization problem at small computational cost\cite{Arsenault_unpublished}. In real materials applications a more complicated system has to be taken into account; Hunds coupling, multi-band etc. However, our presented approach for ML is general enough to be adapted.

L.-F.A. was supported by the Office of Science of the U.S. Department of Energy under Subcontract No. 3F-3138 and A.J.M. by DOE FG-ER04169. L.-F.A. thanks Ara Go for numerous discussions on implementing an exact diagonalization code and Alejandro Lopez-Bezanilla for helpful discussions.


\pagebreak
\begin{center}
\textbf{\large Supplemental Material for Machine learning for many-Body physics: efficient solution of dynamical mean-field theory}
\end{center}

\setcounter{equation}{0}
\setcounter{figure}{0}
\setcounter{table}{0}
\setcounter{page}{1}
\makeatletter

\section{I. Density of states for the simple cubic lattice}
\begin{figure}[!ht]
  \begin{center}
  \includegraphics[scale=0.6]{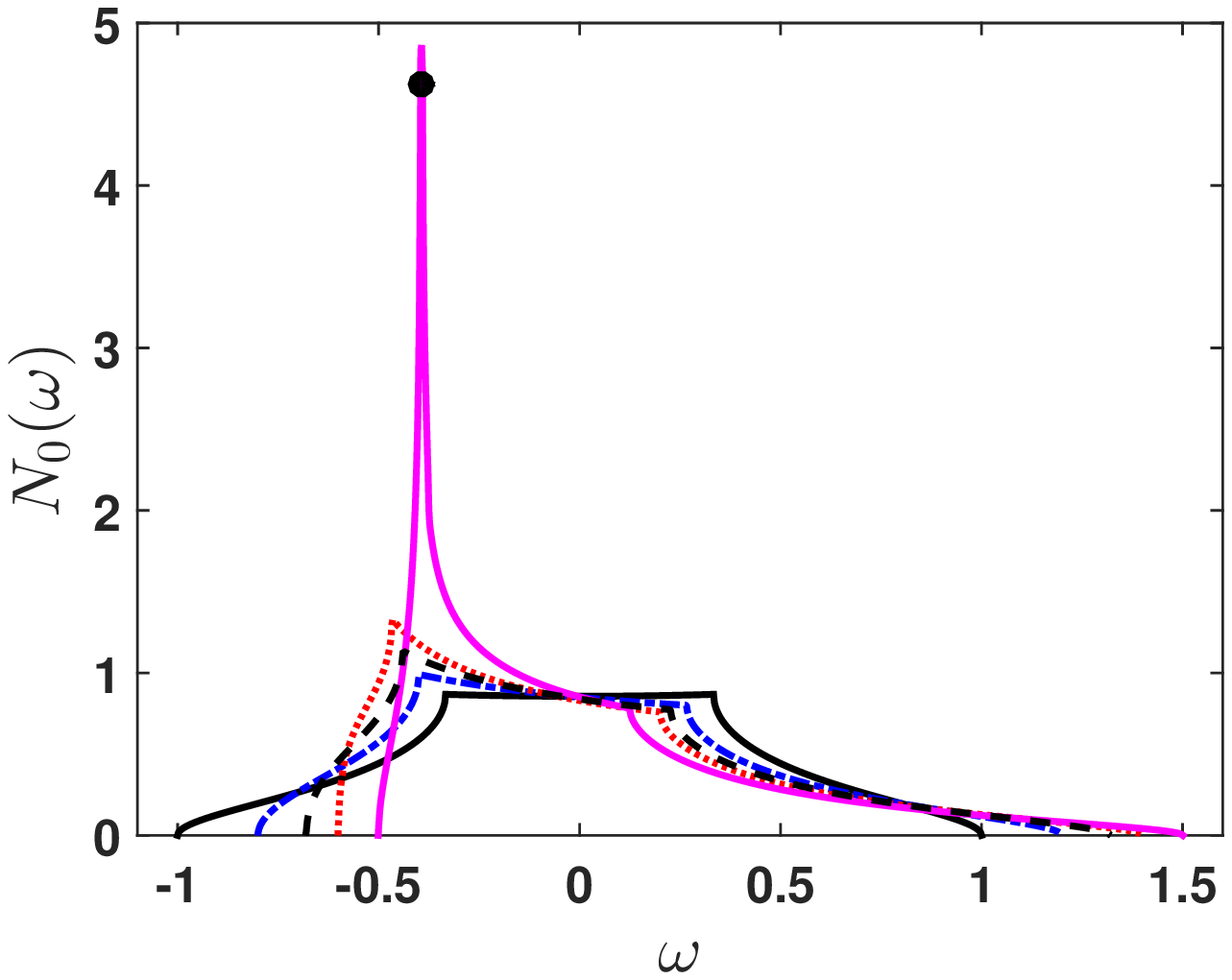}
  \end{center}
  \caption{(Color online) Non-interacting density of states for the three dimensional simple cubic lattice with different values of the next nearest neighbor hopping $t' = 0$ (black solid curve), $t' = -0.1t$ (blue dash-dot curve), $t' = -0.16t$ (black dashed curve), $t' = -0.2t$ (red dot curve) and $t' = -0.3t$ (magenta solid curve and filled black circle).}
  \label{fig:N0}
\end{figure}

The dispersion relation for the the single band tight-binding simple cubic lattice with nearest and next-nearest hopping is given by
\begin{equation}\label{sc_band}
\begin{split}
  \varepsilon_k = -2t\sum_{\alpha=1}^3&\cos\left(k_{\alpha}\right) - 4t'\Big[ \cos\left(k_{1}\right)\cos\left(k_{2}\right)\\ &+ \cos\left(k_{1}\right)\cos\left(k_{3}\right) + \cos\left(k_{2}\right)\cos\left(k_{3}\right)\Big],
\end{split}
\end{equation}
where $-\pi \leq k_{\alpha} \leq \pi$ $\alpha=1,2,3$ labels the three Cartesian directions of the nearest neighbor bonds of the cubic lattice. The bandwith $W=\varepsilon_{[\pi,\pi,\pi] \text{ or } [0,\pi,\pi]}-\varepsilon_{[0,0,0]}$ is given by
\begin{equation}\label{width}
  W =
  \begin{cases}
    12t & \mbox{for }|t'|\leq t/4\;,\\
    8t+16|t'| & \mbox{for }|t'|>t/4\;.\\
  \end{cases}
\end{equation}
As mentioned in the main text, we define the energy unit by fixing $W = 2$. This fixes the value of $t$ for the different $t'$. Our database contains data for $t' = [0,-0.1t,-0.2t,-0.3t]$. We also tested our predictive power by using a lattice with $t'=-0.16t$. We show in Fig.~\ref{fig:N0}. the density of states
\begin{equation}
N_0(\omega)=\sum_k\delta\left(\omega-\varepsilon_k\right)
\label{dosdef}
\end{equation}
of these five lattices to show the effect of next neighbor hopping.
\section{II. Details of the Exact Diagonalization Solver}
\subsection{A. Fitting of the non-interacting hybridization function}
In exact diagonalization, the bath is replaced by a finite number of sites ($N_b$), each characterized by an onsite energy $\varepsilon_l$ and hybridization with the impurity $V_l$. Therefore, the hybridization function is given by
\begin{equation}\label{hyb_ED}
  \Delta_{ED}(i\omega_n) \equiv \sum_{l = 1}^{N_b}\d{V_l^2}{i\omega_n - \varepsilon_l} \Rightarrow \{\varepsilon_l,V_l\},
\end{equation}
which represents in real frequency the approximation of replacing a continuous function by a sum of poles and strengths. These poles and strengths $\{\varepsilon_l,V_l\}$ have to be found by fitting Eq.~\eqref{hyb_ED} to the target function $G_{imp,0}$ obtained via the lattice Green's function $G_{Lattice}^{-1}(i\omega_n) = G_{imp,0}^{-1}(i\omega_n) - \Sigma(i\omega_n)$. This is achieved by defining a distance function and minimizing it.
\begin{equation}\label{distance_appen}
  d = \d{1}{N+1}\sum_{n=0}^{N}W(\omega_n)\left|G_{imp,0}^{-1}(i\omega_n)-G_{imp,0}^{-1,Ns}(i\omega_n)\right|^2,
\end{equation}
where the function $W(\omega_n)$ is chosen to give more weight to some frequencies if wanted. We use $W(\omega_n) = \d{1}{\omega_n}$ to have a better fit of the low frequencies. $N$ is the maximum number of frequency used to define $d$. What is important in its choice is that $\omega_{N_{max}} \gg max (\varepsilon_l)$. Finally, $G_{imp,0}^{-1,Ns}(i\omega_n)$ is the inverse non-interacting Green's function of the Hamiltonian with a finite number of bath sites and is written as
\begin{equation}\label{G0_ED_appen}
  G_{imp,0}^{-1,Ns}(i\omega_n) = z + \mu - \underbrace{\sum_{l=1}^{N_b}\d{V_l^2}{z-\varepsilon_l}}_{= \Delta_{ED}}.
\end{equation}
Therefore, we need to find the set of parameters $\{\varepsilon_l,V_l\}$ that minimizes $d$. This is a problem of unconstrained optimization in several variables. In DMFT, this is done as many time as necessary to converge the solution. We can define a bare hybridization function, for $U = 0$, a function containing the information about the crystal structure and chemistry of the problem. For simplicity, we choose to fix $\mu = 0$. For the non-interacting case, $\Sigma(i\omega_n) = 0$ and the lattice Green's function is given by the band dispersion only
\begin{equation}\label{G0_latt}
  G^0_{Lattice}(i\omega_n) = \sum_{k}\d{1}{i\omega_n - \varepsilon_{k}}.
\end{equation}
We can therefore define the bare hybridization function
\begin{equation}\label{hyb_0}
  \Delta^0(i\omega_n) = i\omega_n - \left(\sum_{k}\d{1}{i\omega_n - \varepsilon_{k}}\right)^{-1}.
\end{equation}
This can be fitted to $\{\varepsilon_l^0,V_l^0\}$ using Eq.~\eqref{distance_appen} and gives a representation of the crystal of size equal to $2N_b$ where $N_b$ is the number of bath sites. The function in this case is thus represented in this basis as $\Delta^0(i\omega_n) \rightarrow \boldsymbol{f} = (\varepsilon_1^0,\varepsilon_2^0,\dots,\varepsilon_{N_b}^0,V_1^0,V_2^0,\dots,V_{N_b}^0)$. It could perhaps be seen as a general and compact way to describe the lattice for ML, irrespective of how the database is constructed (using ED or not). After all, what is needed for creating a $\mathbf{D}$ is a unique way to describe a problem and $[\{\varepsilon_l^0,V_l^0\},U,\mu]$ provides one. This is true for both for model Hamiltonians and real materials as well as for solutions obtained from ED, quantum Monte Carlo etc.
\subsection{B. Particularities of the insulating state in ED}
The database contains 1783 converged DMFT standard metal solutions, 218 converged critical metal solutions and 494 Mott insulating solutions, obtained from exact diagonalization solutions of the DMFT equations. In the Mott insulating state, physically, any choice of chemical potential in the gap  should lead to the same solution with a shifted zero point of the  frequency axis. However, the bath discretization of the ED method means that for otherwise identical parameters different values of $\mu$  lead to  different insulating solutions. For this reason we need to include many (here 494) different Mott insulators in the database. We train insulators using all these solutions.
\section{III. Cross Validation}
\subsection{A. Principle and process}\label{cross_princ}
\begin{figure}[htpb!]
  \begin{center}
  \includegraphics[scale=0.5]{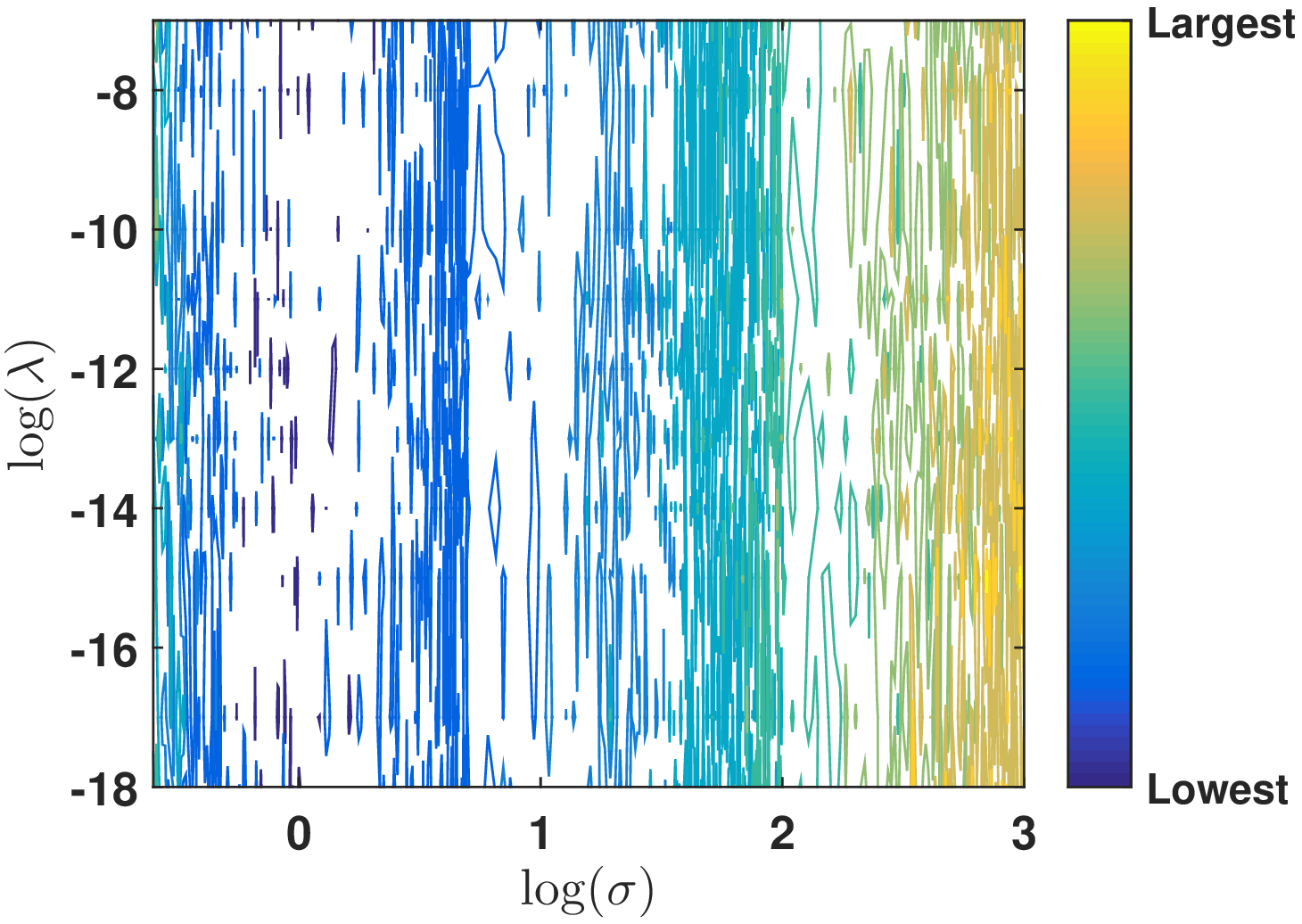}
  \end{center}
  \caption{(Color online) Contour plot showing the mean absolute error for the first five Legendre polynomials coefficients as a function of different hyperparameters pair $[\sigma,\lambda]$ in cross-validation for the impurity Green's function when the descriptor is chosen to be $\mathbf{D} = [\{\varepsilon_l^0,V_l^0\},U,\mu]$.}
  \label{fig:cross_val}
\end{figure}

\begin{figure}[!ht]
  \begin{center}
  \includegraphics[scale=0.5]{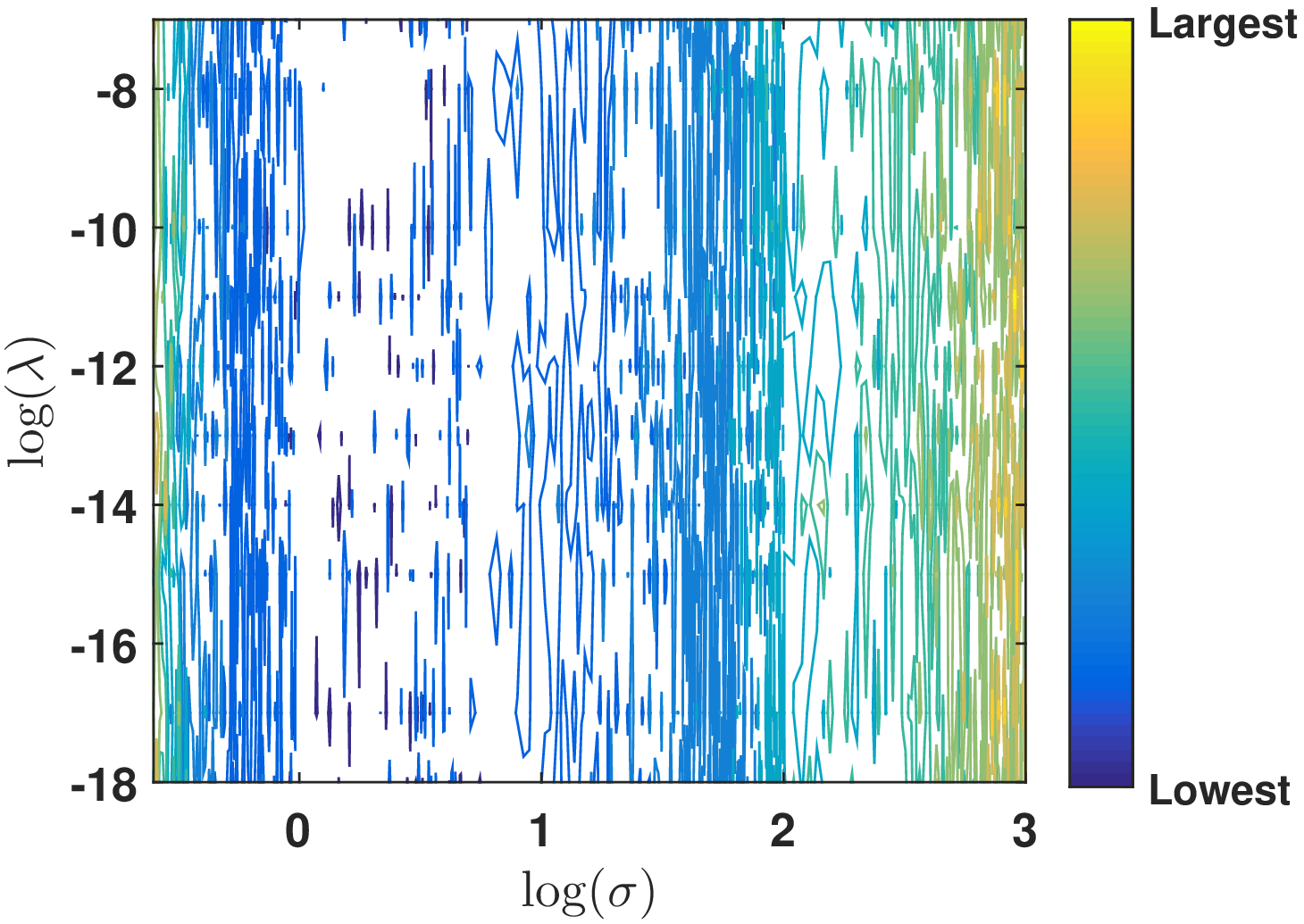}
  \end{center}
  \caption{(Color online) Contour plot showing the mean absolute error for the first five Legendre polynomials coefficients as a function of different hyperparameters pair $[\sigma,\lambda]$ in cross-validation for the impurity Green's function when the descriptor is chosen to be $\mathbf{D} = [\{\Delta_l^0\},U,\mu]$.}
  \label{fig:cross_val_Legendre}
\end{figure}

In kernel ridge regression (KRR), there are two free parameters or hyperparameters. In our case, for $K$, we use the weighted exponential kernel
 \begin{equation}\label{Kernel_def}
    K(\mathbf{D}_i,\mathbf{D}) = \text{e}^{-\d{|\boldsymbol{d}_i|}{\sigma}},
  \end{equation}
where $|\boldsymbol{d}_l| = |\text{D}_{l1}-\text{D}_{1}|+ |\text{D}_{l2}-\text{D}_{2}| + \ldots$ is the Manhattan distance between the two parameter sets in descriptors space and $\sigma$ gives the radius of effect that a particular point of the data set $\mathbf{D}_l$ will have in the prediction process. Therefore, the two free parameters are $\sigma$ (entering the kernel function) and $\lambda$, the regularization parameter used in the cost function that is minimized in KRR. To fix them, we use cross-validation. Cross validation proceeds by first creating a large number of pairs $[\sigma,\lambda]$. Then, for each pair we randomly split the database so that the testing set contains about ten examples. For each of these ($\sim 10$) tests, we predict only the first five Legendre polynomials coefficients $G_{l=0\ldots 4}$ and calculate the total mean absolute error. Note that the actual metric to calculate the error is not really important, we are just looking for the pair that minimizes an error. We look for a pair $[\sigma,\lambda]$ that gives as small as possible error, then this $[\sigma,\lambda]$ is used to learn all the necessary Legendre polynomials coefficients and not only the first five. As an example, we show the result for the training of the metallic impurity Green's function as a contour plot in Fig.~\ref{fig:cross_val} when the descriptor uses ED representation for the non-interacting hybridization function and Fig.~\ref{fig:cross_val_Legendre} when the Legendre representation for the non-interacting hybridization function is chosen. The white color indicates regions where no data are available. We see that if the width $\sigma$ of the kernel is too small (ML will use not enough solutions in descriptors spaces) or too large (ML will use too many far away solutions in descriptors spaces), the error is the largest. For the training/testing using solved problems from our database, the best possible $\sigma$ would be around but smaller than $\sigma = 1$ for Fig.~\ref{fig:cross_val} and around $\sigma = 1.5$ for Fig.~\ref{fig:cross_val_Legendre}. The results also show that the value of $\lambda$ is not extremely important. Practically we chose the pair $[\sigma,\lambda]$ that produced the smallest error among the ones tested while we increase the $\sigma$ to prevent overfitting in the case of out of database prediction(see  Section~III.B below).
\subsection{B. Overfitting and prediction for a new problem}\label{overfit}
\begin{figure}[htpb!]
  \begin{center}
  \includegraphics[scale=0.5]{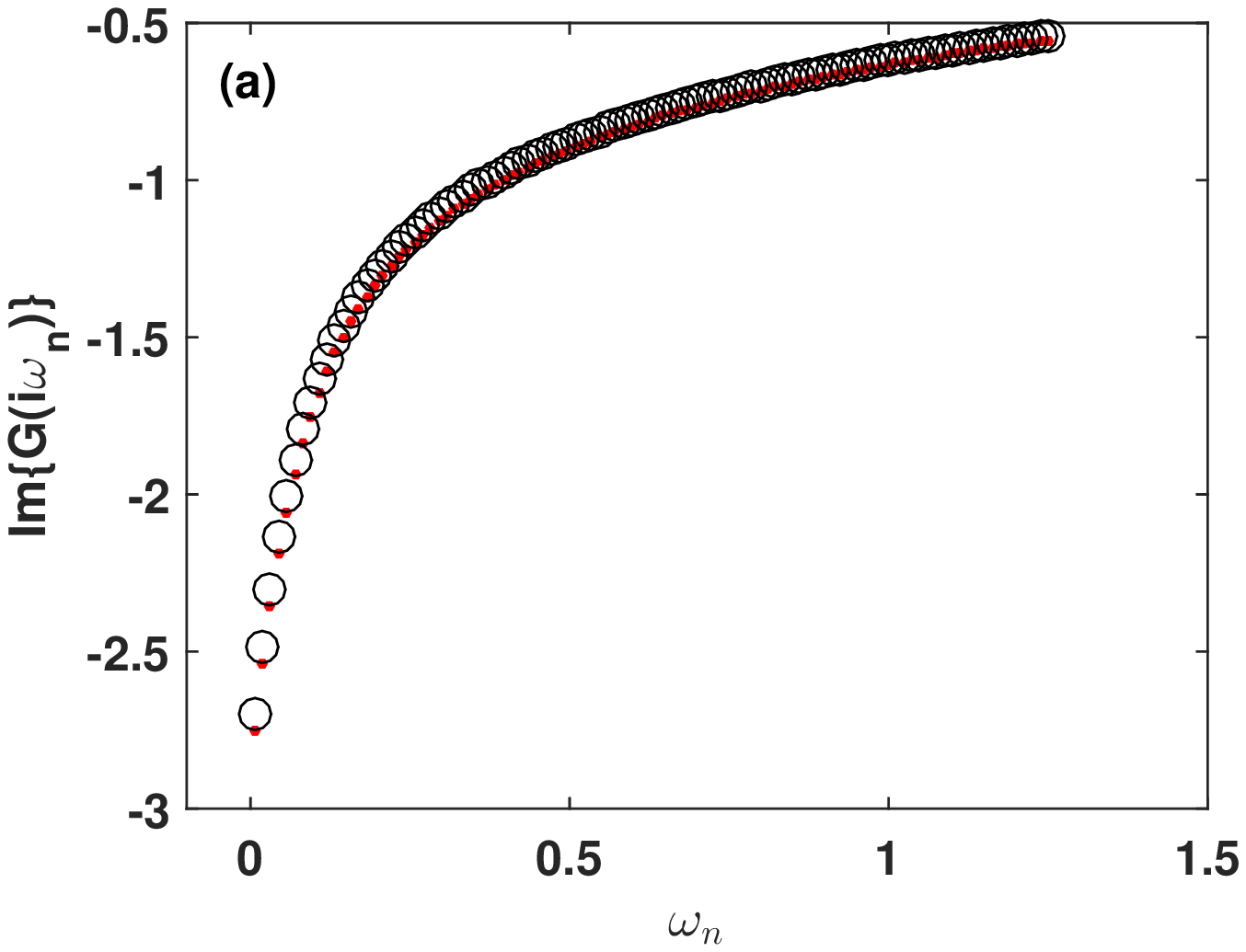}
  \includegraphics[scale=0.5]{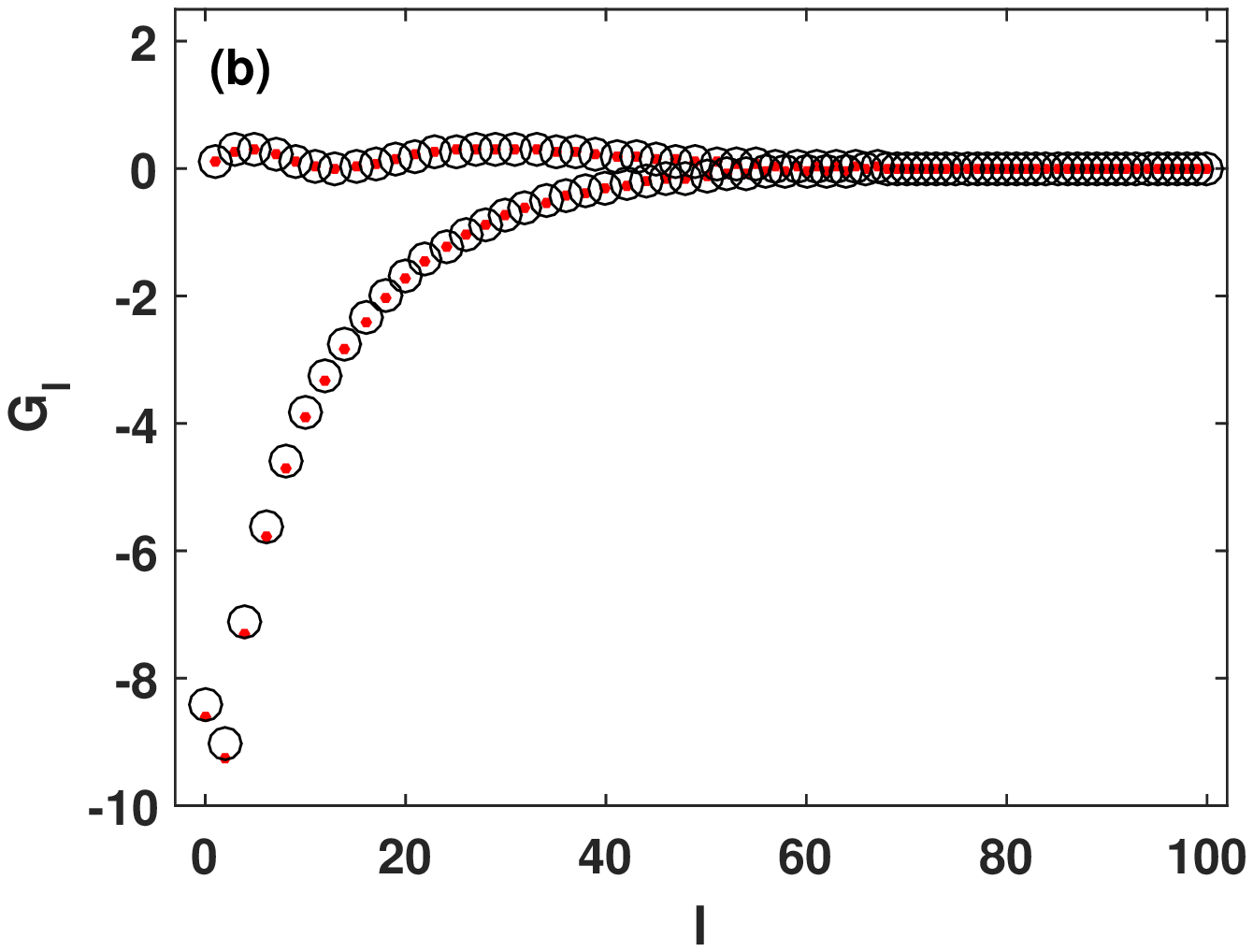}
  \includegraphics[scale=0.5]{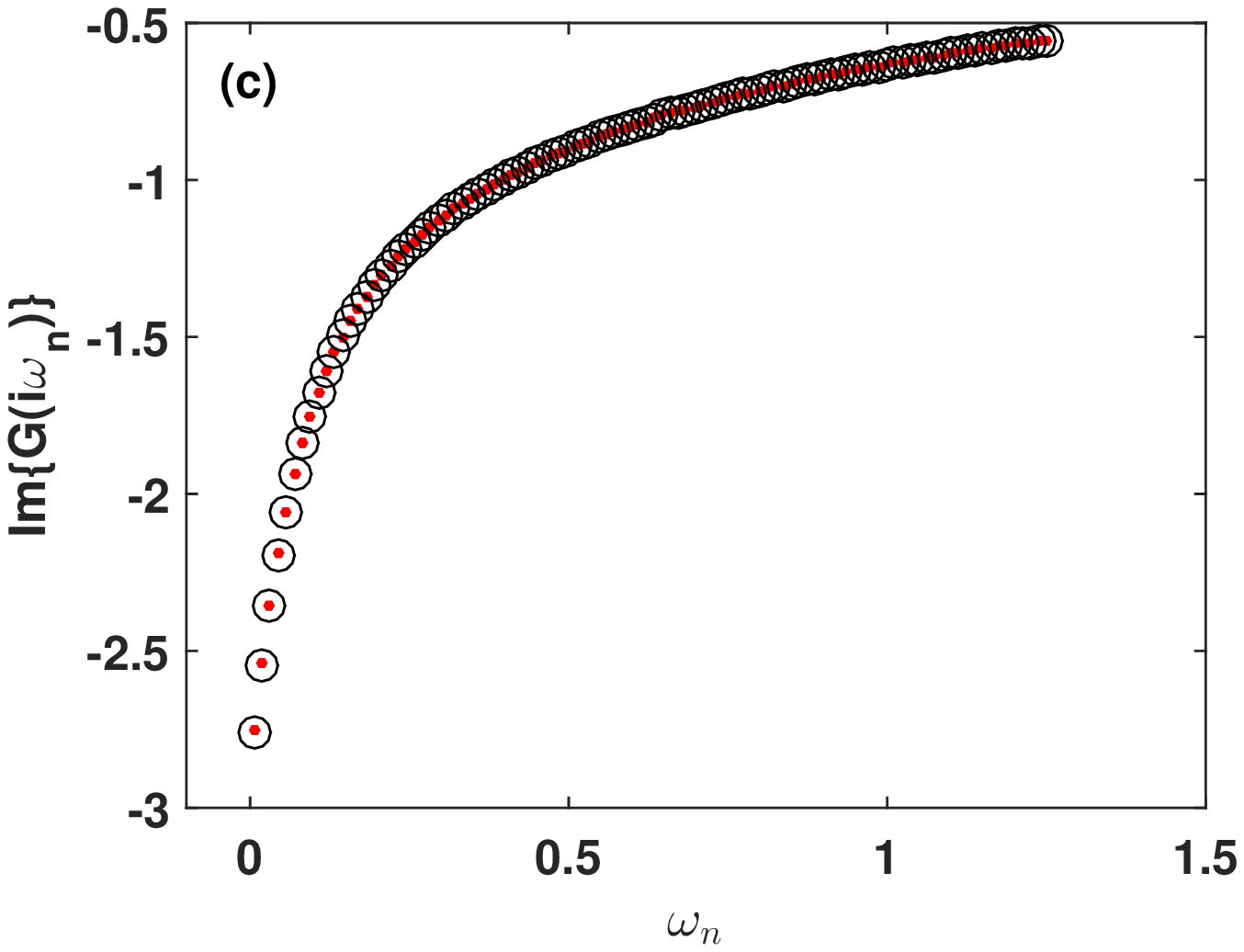}
  \end{center}
  \caption{(Color online) Machine Learning predicted impurity Green's function (black circles) with descriptor $\mathbf{D} = [\{\varepsilon_l^0,V_l^0\},U,\mu]$ as compared to the exact results (red dots) for $U = 2$, $t' = -0.16t$, $n_d \approx 0.85$ (a) Prediction of $\text{Im}\{G(i\omega_n)\}$ with a too small (over-fitted) width of kernel (b) Prediction of $G_l$ with a too small (over-fitted) width of kernel (c) Prediction of $\text{Im}\{G(i\omega_n)\}$  with a right width of kernel.}
  \label{fig:ML_G_predic_outside_supp}
\end{figure}

Here we discuss in detail the prediction of the DMFT solution for the case $t' = -0.16t$, $U = 2$ and $n_d \sim 0.85$ which is a case completely outside our database. Let us first use the width of the kernel that gave the lowest error during cross-validation using $\mathbf{D} = [\{\varepsilon_l^0,V_l^0\},U,\mu]$ as the descriptor which is around $\sigma = 1$ as shown is Section~III.A. The result for the imaginary part of the reconstructed impurity Green's function is shown in Fig.~\ref{fig:ML_G_predic_outside_supp}-(a) while Fig.~\ref{fig:ML_G_predic_outside_supp}-(b) shows the ML predicted coefficients $G_l$. We see a systematic discrepancy in the prediction of the reconstructed $\text{Im}\{G(i\omega_n)\}$, but it is interesting to realize that this is due to a systematic discrepancy of $\sim 2\%$ only on the even $G_l$ coefficients at low order. This is true for every possible example at $t' = -0.16t$ (we calculated a total of 295 different examples with this $t'$). The discrepancy is not due to a failing of ML but solely on a too tight choice of kernel width (over-fitting) which for example might not use enough of the solutions for $t' = -0.1t$ and/or $t' = -0.2t$. We trained our machine again, but with a larger width of $\sigma = 10$ which, according to Fig.~\ref{fig:cross_val} is not the optimal value but still give a pretty acceptable error. The result is Fig.~5 of the main text and is reproduced here as Fig.~\ref{fig:ML_G_predic_outside_supp}-(c) showing that we must be careful with over-fitting and if this is properly taken into account, we can predict new DMFT solutions. The specific choice of $\sigma$ is not well defined since out of database predictions means no comparison with exact results in principle. A sensible approach is to start from the $\sigma$ given by cross-validation of Section~III.A and search for a larger value. To do so, contour plots of Figs.~\ref{fig:cross_val} and \ref{fig:cross_val_Legendre} are essential. We need a value to the right of the cross-validated one since we are looking for a larger radius of effect for database points. How large will be given by the condition of still keeping a low enough error for intra-database prediction. Hence by looking at Figs.~\ref{fig:cross_val} and \ref{fig:cross_val_Legendre} we see that values of $\sigma$ somewhere between $\log(\sigma) = 0.5$ and $\log(\sigma) = 1$ respect these conditions. We chose $\log(\sigma) = 1$, but we could have tested many other values.
\section{IV. Obtaining the Legendre coefficients}
We expand the output solution in terms of Legendre polynomials. Considering the result as a function of imaginary time $0<\tau<\beta$ ($\beta=1/T$ is the inverse temperature) we have for correlation functions \cite{Boehnke_Legendre_supp}
\begin{equation}
f(\tau) = \sum_{l=0}^{\infty}\d{\sqrt{2l+1}}{\beta}f_lP_l(x(\tau)).
\label{Legendredef}
\end{equation}

The ED method we used to do DMFT calculations provides correlation functions in energy space (real or Matsubara frequency); we must therefore Fourier transform the result. To perform the Fourier transform we note that the functions $\Delta(i\omega_n)$, $G_{imp}(i\omega_n)$, $G_{Lattice}(i\omega_n)$ and $\Sigma(i\omega_n)$ of interest here all decay as $C/i\omega_n$ at large $|\omega_n|$. Therefore, let us define the general Matsubara frequency function $f(i\omega_n)$ with asymptotic behaviour $f(i\omega_n)_{n\rightarrow\infty} \rightarrow \d{C}{i\omega_n}$. Depending on the specific correlation function, the constant $C$ is given by
\begin{equation}\label{C_asymp}
  C =
  \begin{cases}
    1 & \mbox{for } G\;,\\
    U^2\d{n_d}{2}\left(1-\d{n_d}{2}\right) & \mbox{for } \Sigma\;,\\
    \sum_k\varepsilon_k^2 \underbrace{=}_{\text{ED case}} \sum_lV_l^2 & \mbox{for } \Delta\;.\\
  \end{cases}
\end{equation}
The imaginary time function is given by Fourier transform $f(\tau) = T\sum_{n}\text{e}^{-i\omega_n\tau}f(i\omega_n)$. Direct numerical evaluation of the sum is complicated by the slow $1/\omega_n$ decay.  We therefore treat the $1/\omega_n$ term analytically:
\begin{eqnarray}\label{h_tau}
f(\tau) & = & T\sum_n \text{e}^{-i\omega_n\tau}\left[f(i\omega_n)- \d{C}{i\omega_n} + \d{C}{i\omega_n}\right]\\
& = & 2T\sum_{n>0}\Big[ \text{Re}\{f(i\omega_n)\}\cos(\omega_n\tau)\nonumber \\
&& ~~~~~~~~~+\: \left( \text{Im}\{f(i\omega_n)\} + \d{C}{i\omega_n}  \right)\sin(\omega_n\tau) \Big]\nonumber \\ && ~~~~~~~~~-\: \d{C}{2}\nonumber .
\end{eqnarray}
This way, we can obtain the value of $f(\tau)$ for any desired $\tau$. Therefore, we can compute the coefficients of the expansion in Legendre polynomials as was done in [\onlinecite{Arsenault2014_supp}] by using the algorithm based on Chebyshev-Legendre transform [\onlinecite{Hale2014}] exploiting the idea that smooth functions can also be represented by expansions in Chebyshev polynomials using fast Fourier transform. This algorithm is implemented in a free MATLAB toolbox called CHEBFUN [\onlinecite{Chebfun}].

Once the coefficients have been obtained, not only the function in imaginary time can be reconstructed\eqref{Legendredef}, but the Fourier transform to $f(i\omega_n)$ can also be done analytically\cite{Boehnke_Legendre_supp} to give
\begin{equation}\label{Leg_exp_fiwn}
  f(i\omega_n) = \sum_{l=0}^{\infty} T_{nl}f_l,
\end{equation}
where $T_{nl} = (-1)^{n}i^{l+1}\sqrt{2l+1}j_l\left(\d{(2n+1)\pi}{2}\right)$ and $j_l(z)$ are the spherical Bessel functions.
\section{V. Choice of representation for the bare hybridization function $\Delta^0(i\omega_n)$}
\begin{figure}[]
  \begin{center}
  \mbox{\includegraphics[scale=0.5]{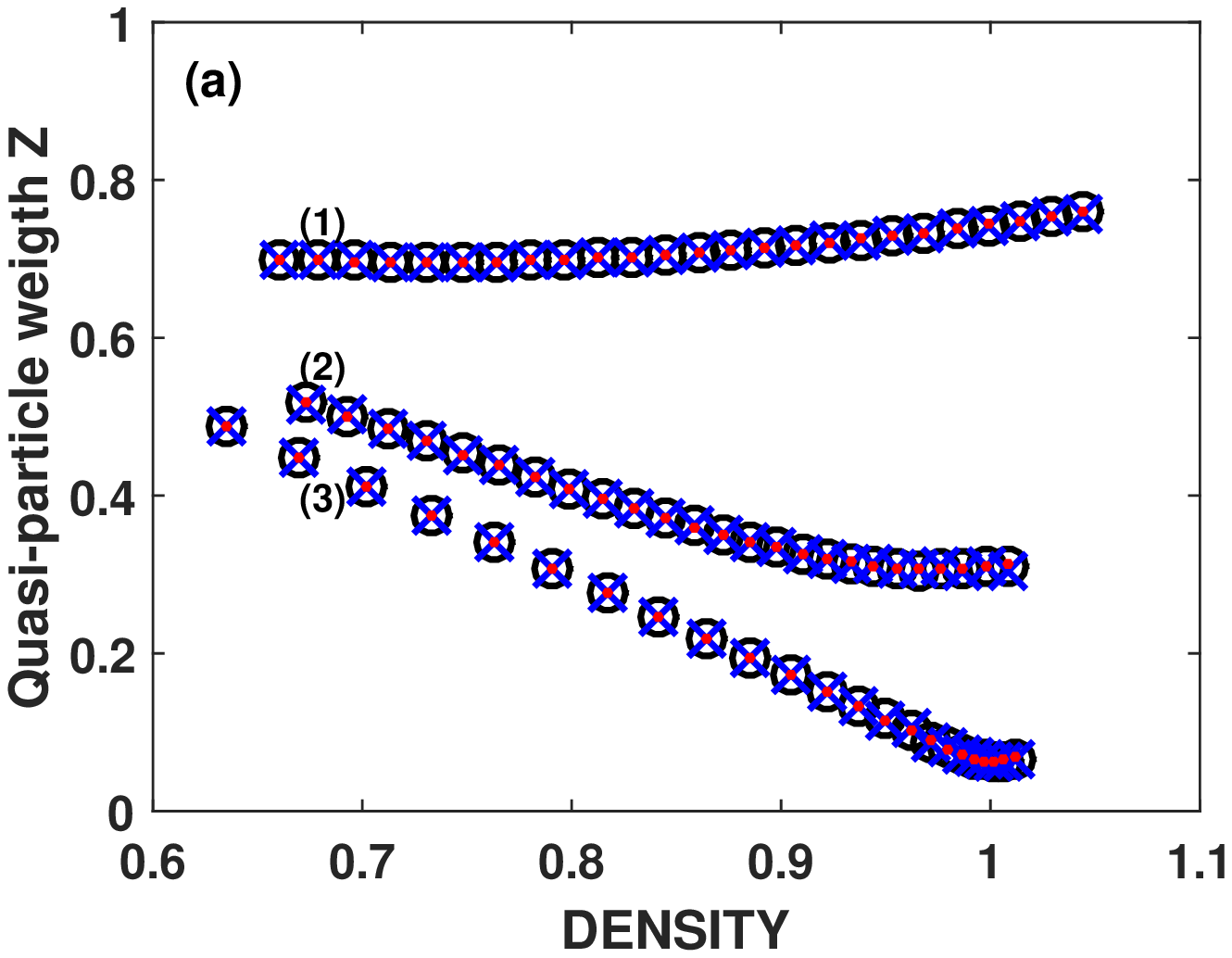}}
  \mbox{\includegraphics[scale=0.5]{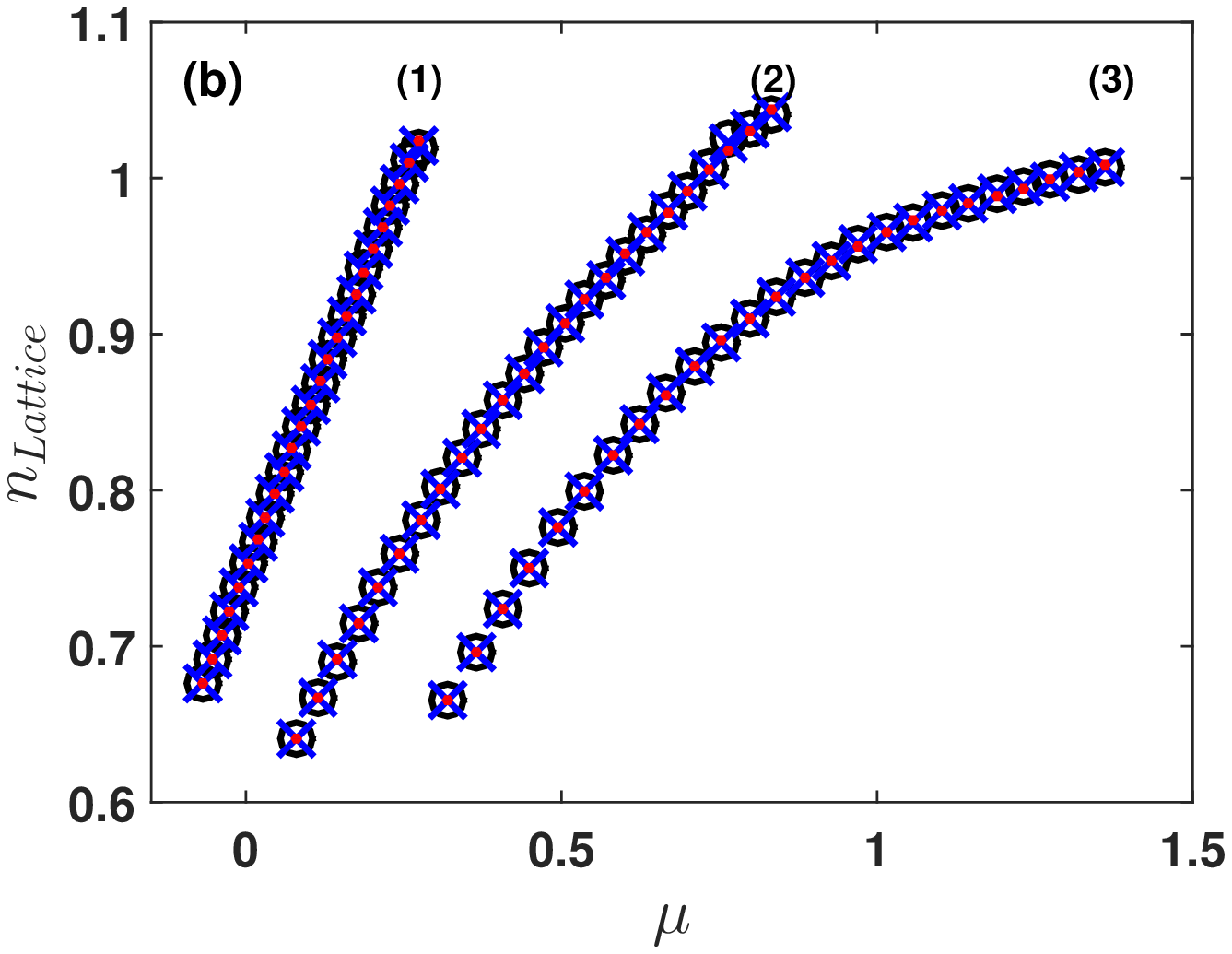}}
  \end{center}
  \caption{(Color online) Machine Learning predictions (black circles) for descriptor $\mathbf{D} = [\{\varepsilon_l^0,V_l^0\},U,\mu]$ and (blue x) for descriptor $\mathbf{D} = [\{\Delta_l^0\},U,\mu]$  as compared to the exact results (red dots). (a) Quasi-particle weight $Z$ as a function of filling of the impurity for different $U$ and $t'$  (1) $U = 0.64$ $t' = -0.3t$, (2) $U = 1.44$ $t' = -0.1t$, (3) $U = 2.08$ $t' = 0$. (b) Lattice density as a function of the chemical potential ($\mu$) for different $U$ and $t'$ (1) $U = 0.64$ $t' = -0.2t$, (2) $U = 1.44$ $t' = 0$, (3) $U = 2.08$ $t' = -0.1t$ (axis shifted $\mu+0.2$).}
  \label{fig:Z_ML_supp}
\end{figure}
In the main text we claimed that the results we presented hold irrespective if the chosen representation of the bare hybridization function is from an ED-like fitting $\{\varepsilon_l^0,V_l^0\}$ or as in term of Legendre polynomials $\{\Delta_l^0\}$ expansion. We show here two examples to support our statement. In Fig.~\ref{fig:Z_ML_supp}-(a) and (b), we reproduce Fig.~2-(a) and Fig.~3-(a) of the main text. In addition, we add the predictions (blue x) as obtained from the descriptor of the form $\mathbf{D} = [\{\Delta_l^0\},U,\mu]$. The predictions for the quasi-particle weight are very close to the case when the machine is trained with $\mathbf{D} = [\{\varepsilon_l^0,V_l^0\},U,\mu]$. Other optimizations could be done to get even closer to the exact solutions and thus for intended purpose it can be considered as being equal. The case of the lattice density is similar, the two representations give again practically the same answer. In conclusion, the two choices of representation for the bare hybridization function in the descriptor give essentially the same results and therefore its choice is a matter of which one can be obtained and how easy to calculate it is for a particular situation.
\section{VI. Median relative difference}
We present exactly how we calculated the median relative difference for the quasi-particle weight. The same approach is taken for the density. Its meaning is the following: 1) A size $N_{learning}$ of learning set is chosen. 2) From the database, we select randomly one example that will be the testing system. 3) From the remaining examples in the database, we randomly take $N_{learning}$ solutions and train a machine (calculate the $\alpha$ matrix of KRR). 4) We predict the self-energy for the testing example of 2) using our trained machine. From it, we obtain $Z$ and can calculate the relative difference with the exact answer $100\d{\left|Z_{ML}-Z_{exact}\right|}{\left|Z_{exact}\right|}$. We then repeat steps 3) and 4) twenty times to obtain the prediction of the same example from many different trained machines and thus assure we have large distributions of learning sets. Finally, we go back to step 2) and start again with a new example and do it fifty time in total to have a large distribution of testing examples. We therefore have one thousand relative difference. If a randomly chosen learning set only contains examples very far from the testing example we are predicting, the relative difference will be large, but in that case, not because ML is bad, but rather because the learning set was badly picked. Therefore to have a good idea of how well a learning set of size $N_{learning}$ does, we argue that a good choice is to take the median value of the one thousand predictions. For the largest possible size of learning set 1782, we instead calculated the relative difference for every example in the database choosing the reminder 1782 examples as learning set. Then we randomly choose fifty of these relative difference and calculated the median.

\end{document}